\documentclass[aps,prl,twocolumn,longbibliography]{revtex4-2}
\usepackage[utf8]{inputenc}
\usepackage{amsmath}
\usepackage{commath}
\usepackage[pdftex]{graphicx}
\usepackage{siunitx}
\usepackage{hyperref}

\begin{document}
\title{On criticality of interface depinning\\ and origin of "bump" in the avalanche distribution}
\author{Lasse Laurson}
\affiliation{Computational Physics Laboratory, Tampere University, P.O. Box 600, FI-33014 Tampere, Finland}

\begin{abstract}
The depinning transition critical point is manifested as power-law distributed avalanches exhibited by slowly driven elastic interfaces in quenched random media. Here we show that since avalanches with different starting heights relative to the mean interface height or different initial local interface curvatures experience different excess driving forces due to elasticity, avalanches close to the "global" critical point of non-mean field systems can be separated into populations of subcritical, critical and supercritical ones. The asymmetric interface height distribution results in an excess of supercritical avalanches, manifested as a "bump" in the avalanche size distribution cutoff.
\end{abstract}

\maketitle

Driven elastic interfaces in quenched random media, e.g., domain walls in ferromagnets~\cite{zapperi1998dynamics} and ferroelectrics~\cite{paruch2005domain}, contact lines in wetting~\cite{joanny1984model}, dislocations~\cite{zapperi2001depinning} and crack fronts~\cite{laurson2013evolution} in disordered solids, often exhibit critical-like response to external driving forces. Such features emerge due to he interplay between quenched disorder, elasticity, and an external driving force, resulting in a depinning transition between pinned and moving phases of the interface at a critical external force $F_{\mathrm{c}}$ at $T=0$~\cite{nattermann1992dynamics,chauve2000creep}. For $T>0$, thermal rounding of the transition~\cite{bustingorry2007thermal} together with creep motion below $F_{\mathrm{c}}$~\cite{ferrero2021creep} would ensue. A key dynamical signature of depinning criticality is the avalanche dynamics of slowly driven interfaces: they move in a sequence of avalanches of widely varying sizes, with the avalanche size distribution $P(s)$  exhibiting power law scaling with exponents that depend on the universality class of the system~\cite{wiese2022theory}. As an example, for linear 1D elastic interfaces in 2D quenched random medium, one can distinguish three such universality classes which emerge due to local, long-range and infinite range (mean field) elastic interactions, respectively~\cite{tanguy1998individual,laurson2013evolution}. 

In practice, due to the presence of various mechanisms (such as the demagnetising field for domain walls in ferromagnets~\cite{skaugen2019analytical,zapperi1998dynamics}) limiting the avalanche growth, the power law $P(s) \sim s^{-\tau}$, with $\tau$ the avalanche size exponent, does not usually continue up to arbitrarily large avalanche sizes $s$. Instead, the power law terminates around a finite cutoff avalanche size $s^{*}$. The manner in which this happens is encoded into a universal cutoff scaling function $f(x)$, such that $P(s) = s^{-\tau} f(s/s^{*})$. Often it is assumed that $f(x)$ is either a simple exponential (as predicted by mean field theory~\cite{rosso2009avalanche}), or, e.g., a stretched exponential. However, in many systems such a simple form is not applicable as it is empirically observed that a "bump" is present around $s^{*}$~\cite{chevalier2015moving,oyama2021unified,girardi2016griffiths,karimi2017inertia,ruscher2021avalanches,rosso2009avalanche,laurson2010avalanches,janicevic2016interevent}. This implies an excess of large avalanches around $s^{*}$, something that can be observed, e.g., in 
systems where inertial effects play a role~\cite{karimi2017inertia,ruscher2021avalanches}. Importantly, a small but clearly observable bump is known to be present also in simple overdamped models of interface depinning with local elasticity~\cite{rosso2009avalanche} (i.e., the quenched Edwards-Wilkinson (qEW) equation~\cite{kardar1986dynamic,alava2002interface,kim2006depinning}), as well as in the long-range (LR) elastic string~\cite{10.1115/1.3176178,laurson2010avalanches,janicevic2016interevent,bonamy2008crackling}, but it is absent in systems with infinite-range mean field (MF) elasticity~\cite{rosso2009avalanche}. The bump is also predicted by functional renormalisation group (FRG) calculations~\cite{rosso2009avalanche,le2009statistics,le2009size}, but the complexity of FRG does not allow for a straightforward physical understanding of its origin, which therefore remains an important open question.

Here we show that for driven elastic interfaces in random media with local and long-range elasticity, avalanches with different starting heights relative to the mean interface height or those with different local curvatures of the interface around the avalanche initiation point experience different average excess driving forces due to elasticity. Hence, avalanches at the "global" depinning critical point can be separated into populations of subcritical, critical and supercritical ones. Due to an asymmetry in the local height distribution at the depinning threshold~\cite{toivonen2022asymmetric} which here is manifested as a skewed distribution of the avalanche starting heights and local curvatures of the interface, there is an excess of supercritical avalanches, resulting in a bump in the cutoff scaling function $f(x)$. In the MF limit with infinite-range interactions where avalanches are not localised, all large avalanches experience on the average the same elastic force. Hence, the tails of the avalanche distributions are independent of the avalanche starting height, and one recovers an exponential $f(x)$ without a bump. This also suggests that strictly speaking only the MF system is truly critical while finite-dimensional systems exhibit a mixture of behaviours around criticality.

We perform simulations of a discretised version of 1D elastic interfaces with local, long-range and infinite-range elasticity in a 2D quenched random medium.  
The local total force acting on the interface element $i$ located at $x=x_i \equiv i$ (with $i$ an integer from 0 to $L$) along the interface $h(x)=h(x_i) \equiv h_i$ is
\begin{equation}
F(x_i) = F_{\mathrm{el}}(x_i) + \eta(x_i,h_i) + F_\mathrm{ext},
\end{equation}
where the first term on the RHS is either $F_{\mathrm{el,qEW}}(x_i) = \Gamma_0 \nabla^2 h(x_i)=\Gamma_0(h_{i+1}+h_{i-1}-2h_i)$, $F_{\mathrm{el,LR}}(x_i) = \Gamma_0 \sum_{j \neq i} \frac{h_j-h_i}{|x_j-x_i|^2}$, or $F_{\mathrm{el,MF}}(x_i) = \Gamma_0 (\langle h \rangle - h_i)$ for qEW, LR and MF elasticity, respectively (with $\Gamma_0$ the stiffness of the interface), $\eta$ is uncorrelated Gaussian quenched disorder with mean zero and unit variance, and $F_\mathrm{ext}$ is the external driving force~\cite{laurson2013evolution}. The parallel dynamics of the interface is defined in discrete time $t$ by setting the local velocity $v(x_i,t) \equiv h(x_i,t+1)-h(x_i,t) = \theta[F(x_i)]$,
where $\theta$ is the Heaviside step function. We employ quasistatic constant velocity driving which after an initial transient keeps the interface in the immediate proximity of the depinning threshold, such that avalanches are triggered by increasing $F_\mathrm{ext}$ just enough to make exactly one interface element unstable [that is, $F(x_i)>0$ for some $i$] whenever the previous avalanche has ended. This allows us to unambiguously determine the avalanche starting point $(x_{\mathrm{init}},h_{\mathrm{init}})$ which will be crucial for our analysis. During an avalanche, $F_\mathrm{ext}$ is decreased at a rate proportional 
to the instantaneous avalanche velocity, $\dot{F}_\mathrm{ext} = -k/L \sum_i v_i (t)$, where $k$ is a parameter controlling the cutoff of the avalanche size distribution~\cite{laurson2013evolution}. To extract the avalanche size distributions for analysis, we collect data on avalanche sizes $s$ [defined as $s=\sum_{t=0}^T \sum_{i=0}^{L-1}v_i(x_i,t)$, where $T$ is the avalanche duration] from the steady state after the initial transient. Unless stated otherwise, the parameters are set to $L=8192$, $\Gamma_0 = 1.0$, and $k=0.0001$ for the qEW equation, $L=16384$, $\Gamma_0=0.5$, and $k=0.00039062$ for the LR model, and $L=131072$, $\Gamma_0 =0.5$, and $k=0.004$ for the MF model. 
The parameter values are chosen so that the maximum lateral size of the avalanches remains smaller than the system size $L$ in order to avoid finite size effects. For the MF model with zero roughness exponent, this implies tuning $k$ to be such that $s^{*} < L$.

\begin{figure}[t!]
    \centering
    \includegraphics[width=0.5\textwidth]{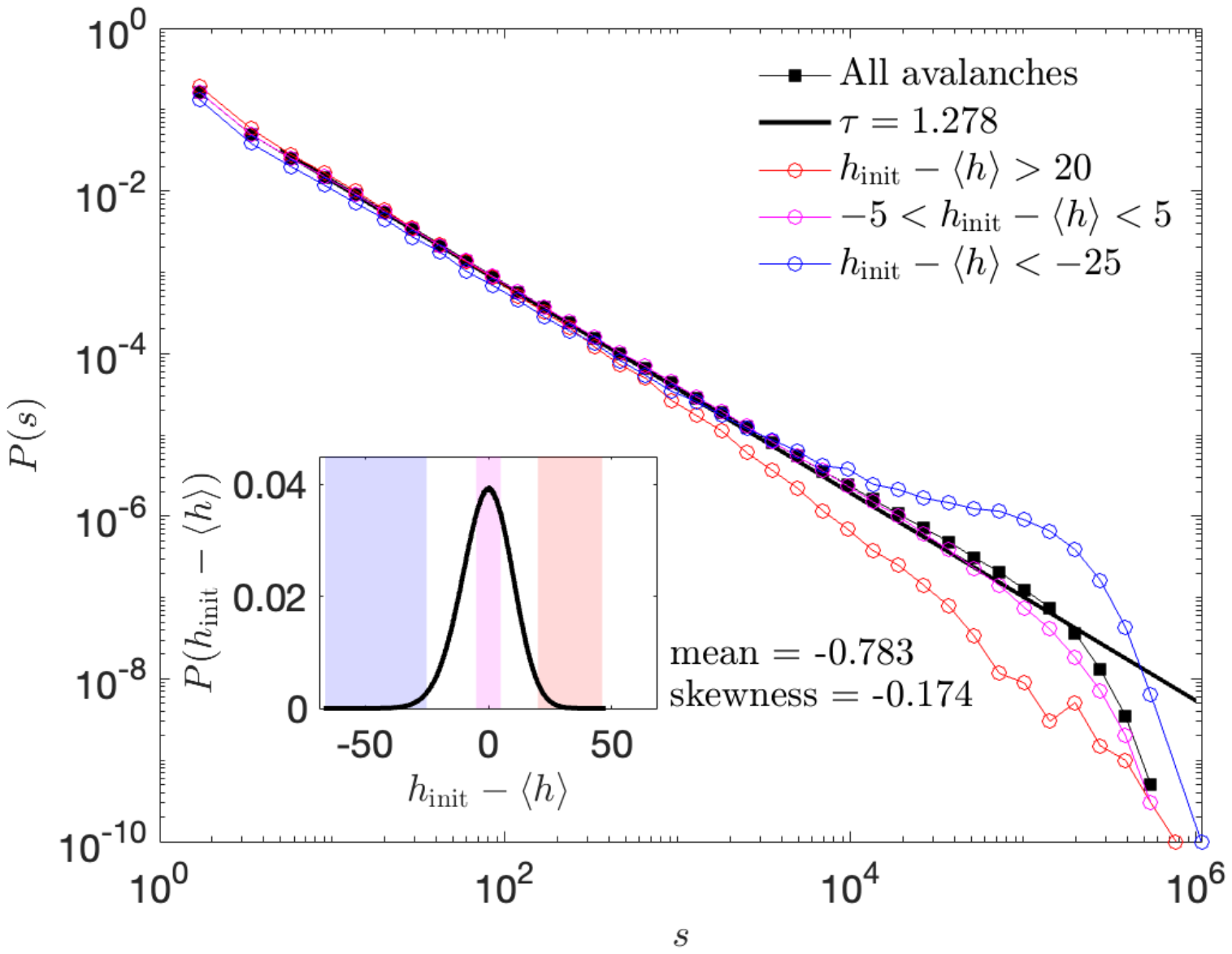}
    \caption{$P(s)$-distributions for the LR model for all avalanches (black) and for different bins of $h_\mathrm{init}-\langle h \rangle$ (colors as in the legend; main figure). The black solid line corresponds to a power law with the exponent $\tau=1.278$. The inset shows the distribution $P(h_\mathrm{init}-\langle h \rangle)$, with the bins used for the $P(s)$-distributions in the main figure shown as shaded regions.}
    \label{fig:1}
\end{figure}

We start by considering the LR model. The steady state $P(s)$ of all avalanches shown with black symbols in Fig.~\ref{fig:1} exhibits a power law part with an exponent $\tau = 1.28 \pm 0.05$~\cite{laurson2010avalanches,bonamy2008crackling,le2021spatial}(solid black line in Fig.~\ref{fig:1}), which is terminated at a cutoff where the scaling function $f(x)$ exhibits a weak bump~\cite{janicevic2016interevent}, visible as data points above the solid black line close to the cutoff in Fig.~\ref{fig:1}. To understand the origin of the bump, we analyze the avalanche distributions separately for different relative starting heights $h_\mathrm{init}-\langle h \rangle$ (with $\langle h \rangle$ the instantaneous spatial average of $h$) of the avalanches. The distribution $P(h_\mathrm{init}-\langle h \rangle)$, shown in the inset of Fig.~\ref{fig:1}, exhibits a slightly negative mean (the avalanches start on the average slightly behind $\langle h\rangle$) and a negative skewness of -0.17, which is very close to the skewness found recently for the distribution of {\it all} interface heights of the same model in Ref.~\cite{toivonen2022asymmetric}. We then consider $P(s)$ separately for avalanches from different parts of $P(h_\mathrm{init}-\langle h \rangle)$: from the positive and negative tails, as well as for a region around $h_\mathrm{init}-\langle h \rangle = 0$, as indicated by the shaded regions in the inset of Fig.~\ref{fig:1}. This analysis reveals the crucial role of $h_\mathrm{init}-\langle h \rangle$ in determining the shape of the tail of $P(s)$. First, $P(s)$ for the region around $h_\mathrm{init}-\langle h \rangle = 0$ (for $-5 < h_\mathrm{init}-\langle h \rangle < 5$, to be specific) hardly exhibits any bump at all, while a very pronounced bump can be observed for avalanches with $h_\mathrm{init}-\langle h \rangle<-25$ in the negative tail of $P(h_\mathrm{init}-\langle h \rangle)$. Avalanches with $h_\mathrm{init}-\langle h \rangle>20$ [positive tail of $P(h_\mathrm{init}-\langle h \rangle)$] exhibit a subcritical-like distribution where the deviation from power law starts at a smaller $s$-value, and no bump is present.

Hence, the above analysis reveals that by conditioning $P(s)$ with $h_\mathrm{init}-\langle h \rangle$ one may divide the avalanches into populations of subcritical, critical and supercritical ones. For the LR model considered here, we attribute this to the excess driving force due to the long-range elastic interactions experienced by the avalanche during its propagation: For $h_\mathrm{init}-\langle h \rangle \ll 0$, i.e., for parts of the interface lagging behind the mean height, the non-local elasticity appears as a positive excess driving force, which together with the global external force $F_{\mathrm{ext}}$ (tuned close to its "global" critical value $F_{\mathrm{c}}$) gives rise to a supercritical total effective driving force, resulting in a pronounced bump in the corresponding $P(s)$. On the other hand, for $h_\mathrm{init}-\langle h \rangle \gg 0$, i.e., for parts of the interface ahead the mean height, elasticity resists avalanche propagation, i.e., the total effective driving force is subcritical. In a sense, true criticality is obtained only in the immediate vicinity of $h_\mathrm{init}-\langle h \rangle = 0$ where the excess driving force due to elasticity will vanish on the average.

\begin{figure}[t!]
    \centering
    \includegraphics[width=0.5\textwidth]{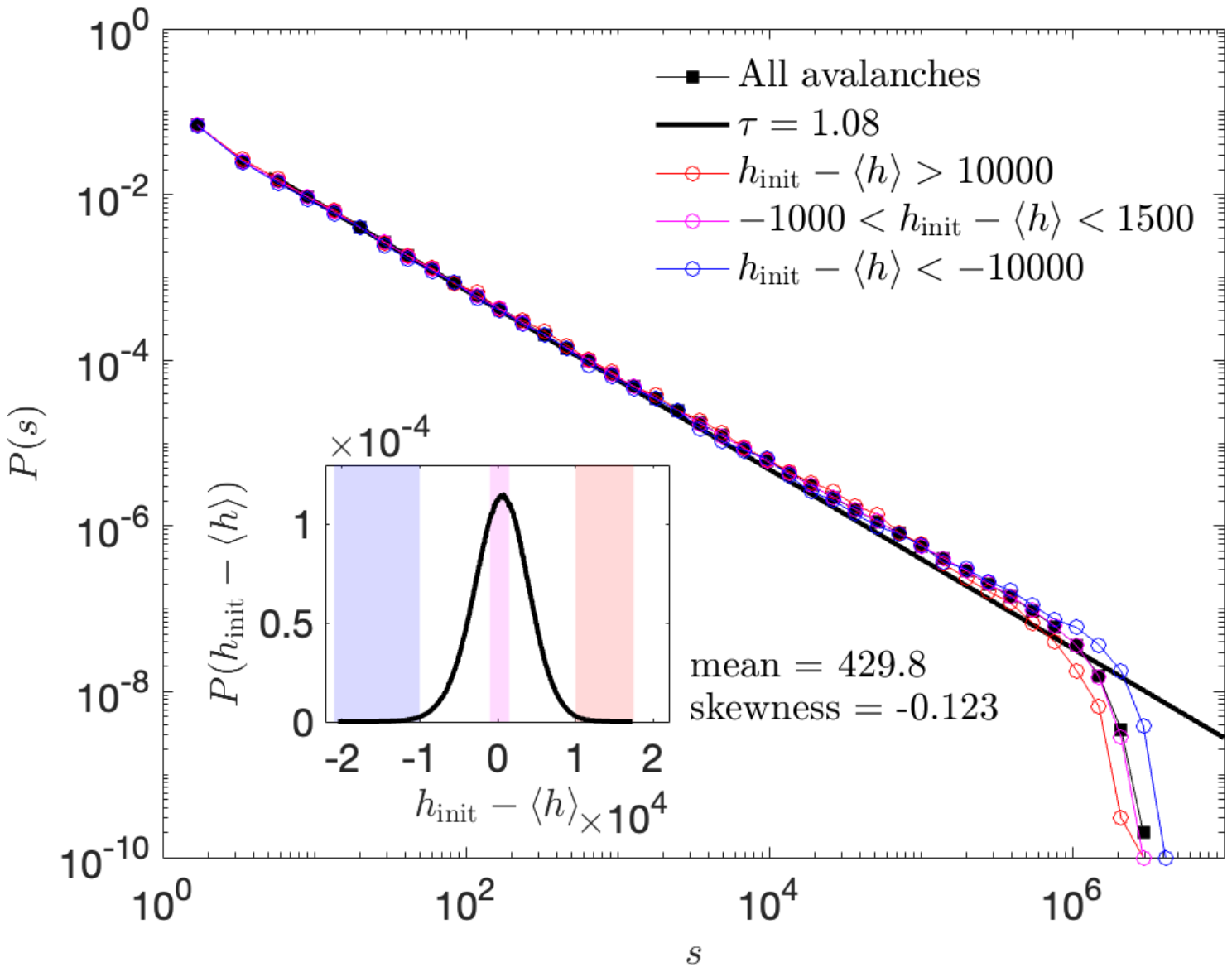}
    \caption{$P(s)$-distributions for the qEW equation for all avalanches (black) and for different bins of $h_\mathrm{init}-\langle h \rangle$ (colors as in the legend; main figure). The black solid line corresponds to a power law with the exponent $\tau=1.08$. The inset shows the distribution $P(h_\mathrm{init}-\langle h \rangle)$, with the bins used for the $P(s)$-distributions in the main figure shown as shaded regions.}
    \label{fig:2}
\end{figure}

\begin{figure}[t!]
    \centering
    \includegraphics[width=0.5\textwidth]{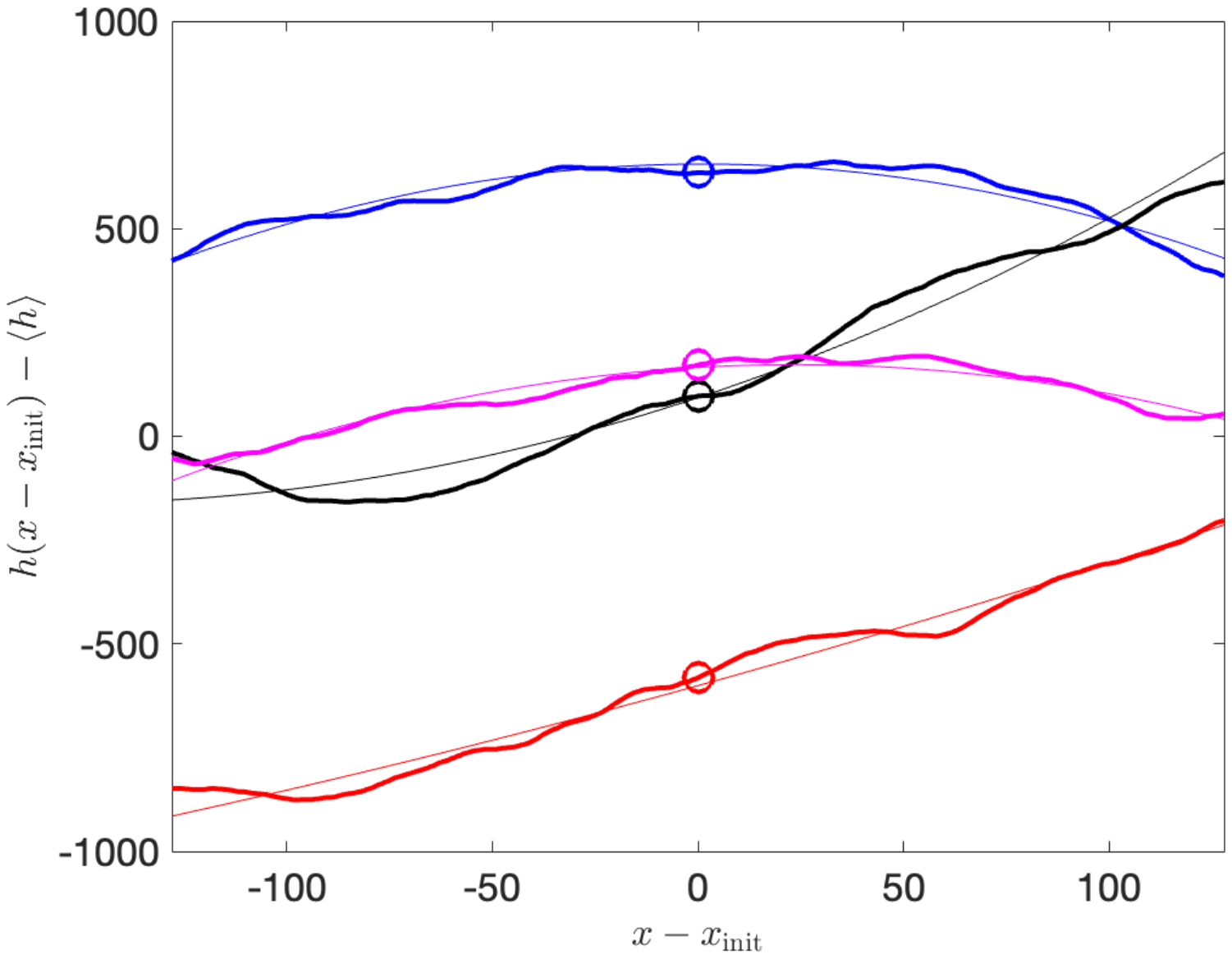}\\
    \includegraphics[width=0.5\textwidth]{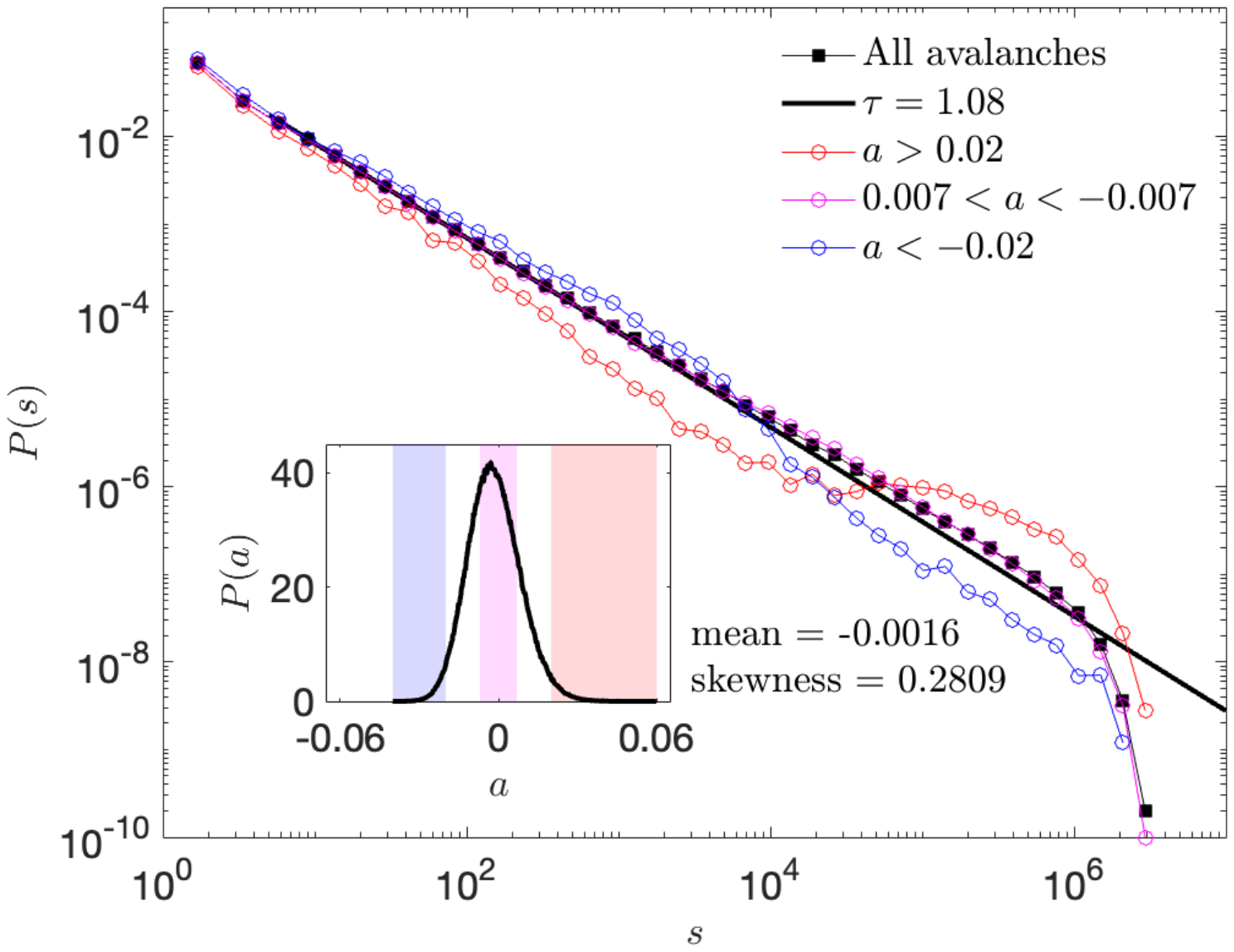}
    \caption{Top: Examples of four interface segments (thick lines) of the qEW equation around $(x_{\mathrm{init}},h_{\mathrm{init}})$, indicated by the circles. The thin lines correspond to parabolic fits to the segments, from which the coefficient $a$ of the quadratic term is taken to characterise the local curvature of the interface in the proximity of $x_{\mathrm{init}}$ (here, within a region of width $l_x = 257$ centered at $x_{\mathrm{init}}$). Bottom: $P(s)$-distributions for all avalanches (black) and for different bins of the $a$-value (colors as in the legend; main figure). The black solid line corresponds to a power law with the exponent $\tau=1.08$. The inset shows the distribution $P(a)$, with the bins used for the $P(s)$-distributions in the main figure shown as shaded regions.}
    \label{fig:3}
\end{figure}

Next, we'll contrast these results with those obtained from a similar analysis of the local qEW equation. Fig.~\ref{fig:2} shows the $P(s)$-distributions for different ranges of  $h_\mathrm{init}-\langle h \rangle$. The difference between $P(s)$ for avalanches from the positive and negative tails of the $P(h_\mathrm{init}-\langle h \rangle)$ distribution [inset of Fig.~\ref{fig:2}; the negative skewness of $P(h_\mathrm{init}-\langle h \rangle)$  of -0.12 is again similar to that found for the distribution of all heights of the same model, see Supplemental Material of Ref.~\cite{toivonen2022asymmetric}] is much less dramatic than in the case of the LR model: apart from a small effect on the maximum avalanche size, the value of $h_\mathrm{init}-\langle h \rangle$ does not play a role for the shape of $P(s)$. This is to be expected as the local nature of elasticity in the qEW equation implies that the elastic force is not strongly dependent on $h_\mathrm{init}-\langle h \rangle$ and hence it is here not the correct quantity to use for conditioning $P(s)$. Thus, to analyse the origin of the bump (which is again visible in Fig.~\ref{fig:2} as data points above the solid line corresponding to a power law with $\tau = 1.08$~\cite{rosso2009avalanche}), we consider instead the effect of the local environment around the avalanche starting point $x_{\mathrm{init}}$, by fitting a parabola $h(x-x_{\mathrm{init}}) = a(x-x_{\mathrm{init}})^2 + b(x-x_{\mathrm{init}}) + c$ to a region of size $l_x$ centered around $x_{\mathrm{init}}$; in the top panel of Fig.~\ref{fig:3}, we consider $l_x = 257$ as an example. The fitting parameter $a$ is then used as a measure of the local curvature of the interface which again implies a positive (for $a \gg 0$) or negative (for $a \ll 0$) excess driving force due to elasticity. The bottom panel shows the distributions $P(s)$ for different ranges of $a$, as illustrated by the shaded areas in the inset, showing the distribution $P(a)$ which exhibits clear positive skewness. Even if here the details are somewhat dependent on the choice of $l_x$, the conclusion is analogous to what we found above for the LR model: avalanches with $a$ in the positive tail of $P(a)$ exhibit a clear bump, while avalanches with $a \ll 0$ have a subcritical character. Hence, considering the effect of the curvature of the local environment of the avalanche starting point allows us to again classify avalanches into subcritical, critical and supercritical ones, with an excess of supercritical avalanches due to the positive skewness of $P(a)$ giving rise to the bump in the full distribution.

\begin{figure}[t!]
    \centering
    \includegraphics[width=0.5\textwidth]{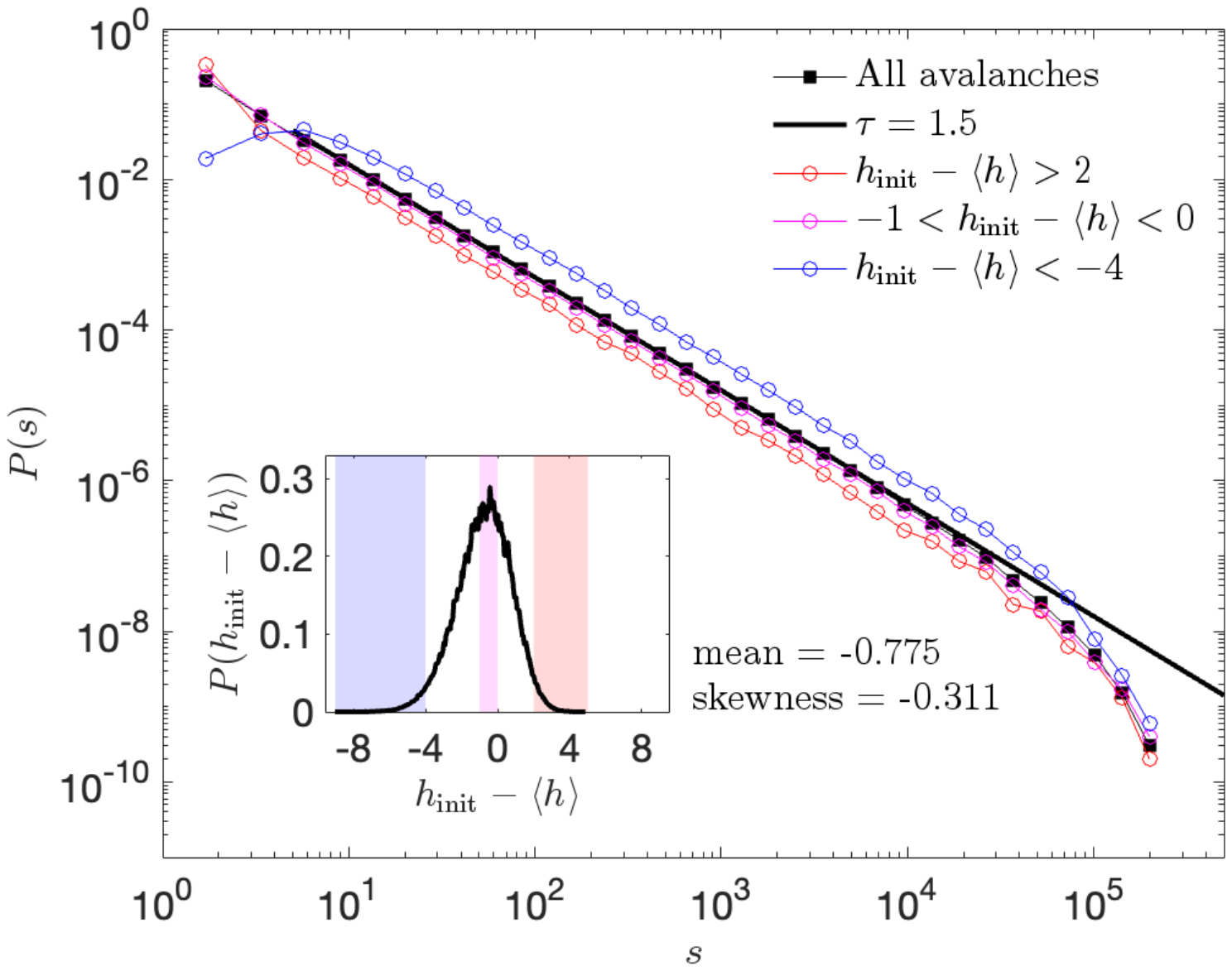}
    \caption{Avalanche size distributions for the MF model for all avalanches (black) and for different bins of the relative avalanche starting height $h_\mathrm{init}-\langle h \rangle$ (colors as in the legend; main figure). The black solid line corresponds to a power law with the exponent $\tau=1.5$. The inset shows the distribution $P(h_\mathrm{init}-\langle h \rangle)$, with the bins used for the $P(s)$-distributions in the main figure shown as shaded regions.}
    \label{fig:4}
\end{figure}

We then consider the MF model, in order to contrast the behaviour it exhibits to the finite-dimensional cases discussed above. Here, we'll again resort to studying the avalanches as a function of $h_\mathrm{init}-\langle h \rangle$, which in this case is equivalent to studying them as a function of $F_{\mathrm{el}}$ acting on $(x_{\mathrm{init}},h_\mathrm{init})$. Fig.~\ref{fig:4} shows the $P(s)$-distributions for different ranges of $h_\mathrm{init}-\langle h \rangle$, as indicated by the shaded regions in the inset; the negative skewness of $P(h_\mathrm{init}-\langle h \rangle)$ is again in agreement with previous results for the distribution of all interface heights for the $\Gamma_0$-value considered here (Supplementary Material of Ref.~\cite{toivonen2022asymmetric}). In this case, the tails of the $P(s)$-distributions do not depend on $h_\mathrm{init}-\langle h \rangle$: In all cases, they are consistent with the known prediction of mean field theory, i.e., a power law terminated at an exponential cutoff, $P(s) \propto s^{-\tau} \exp [-s/(4 s^{*})]$, with $\tau=1.5$~\cite{rosso2009avalanche}. In particular, no bump is present in any of the distributions. The only effect of $h_\mathrm{init}-\langle h \rangle$ concerns the very small avalanches, involving only motion of the triggering site at $x=x_{\mathrm{init}}$, such that avalanches are less likely to be very small (of the order of $s=1$) when $x_{\mathrm{init}}$ is subject to a very large positive elastic force (see the case $h_\mathrm{init}-\langle h \rangle<-15$ in Fig.~\ref{fig:4}), and more likely to have a size of 1 if the elastic force acting on $x_{\mathrm{init}}$ is large and negative (the case $h_\mathrm{init}-\langle h \rangle>-10$ in Fig.~\ref{fig:4}). As soon as the avalanche grows bigger, its starting point and hence the value of $h_\mathrm{init}-\langle h \rangle$ becomes irrelevant as the avalanche is not localised in the MF model. Thus, the tails of the $P(s)$-distributions are independent of  $h_\mathrm{init}-\langle h \rangle$. 

\begin{figure}[t!]
    \centering
    \includegraphics[trim={1.3cm 5.0cm 1.5cm 4.5cm},clip,width=0.5\textwidth]{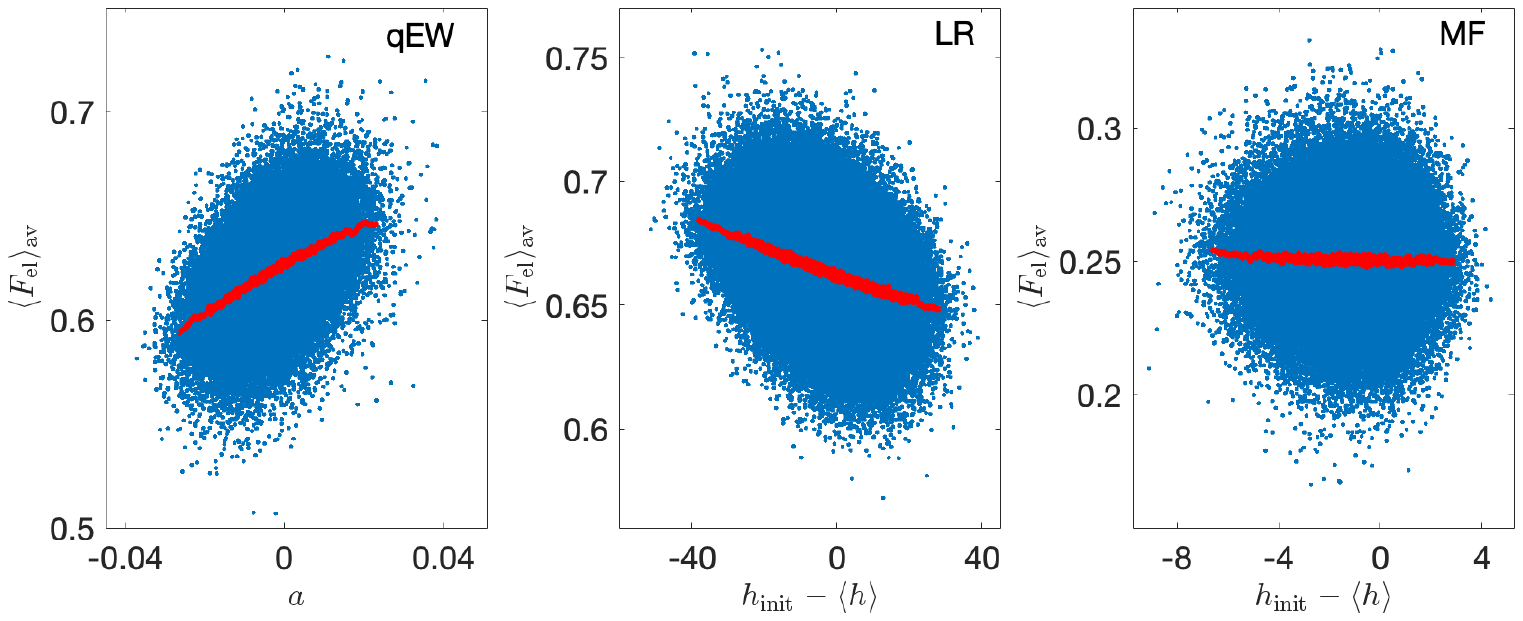}
    \caption{The average elastic force $\langle F_{\mathrm{el}} \rangle_{\mathrm{av}}$ during avalanches (blue dots, $s \in [2000, 20000]$) as a function of $a$ (qEW, left) and $h_\mathrm{init}-\langle h \rangle$ (LR and MF, middle and right, respectively). The running average (red) illustrates how the average of $\langle F_{\mathrm{el}} \rangle_{\mathrm{av}}$ depends on $a$ or $h_\mathrm{init}-\langle h \rangle$.}
    \label{fig:5}
\end{figure}

Finally, we check our assumption behind the discussion up to this point, i.e., how the average $F_{\mathrm{el}}$ depends on $a$ (for the qEW equation) or $h_\mathrm{init}-\langle h \rangle$ (LR and MF models). To this end, we show in Fig.~\ref{fig:5} the average elastic force $\langle F_{\mathrm{el}} \rangle_{\mathrm{av}} = \frac{1}{s}\sum_{t=0}^{T} \sum_{i=0}^{L-1} F_\mathrm{el}(x_i)v(x_i,t)$ for individual avalanches within the scaling regime (considering $s \in [2000, 20000]$; blue dots), together with their running averages (red). In the above expression, $s$ and $T$ are the avalanche size and duration, respectively, and $v(x_i,t)$ multiplies $F_\mathrm{el}(x_i)$ (evaluated before updating the interface position at $x_i$) to make sure that the sum is over active interface segments only.  One can make two main observations: (i) $\langle F_{\mathrm{el}} \rangle_{\mathrm{av}}$ has a positive mean value, which is related to the fact that the average is over interface segments that will move forward during an avalanche (and hence on the average experience a larger $F_{\mathrm{el}}$ than rest of the interface), evaluated before each move, and (ii) there is a clear trend in the running average for qEW and LR, such that avalanches with larger positive $a$ (qEW) and larger negative $h_\mathrm{init}-\langle h \rangle$ (LR) experience on the average a larger $\langle F_{\mathrm{el}} \rangle_{\mathrm{av}}$. In contrast, for the MF model, when considering large enough avalanches such that the weight of the segment at $x=x_{\mathrm{init}}$ becomes negligible, the running average is independent of $h_\mathrm{init}-\langle h \rangle$ and all large avalanches experience on the average the same $\langle F_{\mathrm{el}} \rangle_{\mathrm{av}}$.

What are the implications of this? First, our analysis provides a clear-cut physical explanation for the emergence of the bump in the avalanche cutoff in terms of elasticity that provides on the average an extra contribution to the driving force which depends on either the relative height of the avalanche starting point (LR elasticity) or the local curvature of the interface around the avalanche starting point (local elasticity). For infinite-range elasticity the local elastic force at the avalanche initiation point does not play a role as avalanches are not localised, and hence no bump emerges. Here we have considered $P(s)$ at a fixed average $F_{\mathrm{ext}} \approx F_{\mathrm{c}}$, but the mixture of behaviours around criticality seen here is somewhat similar to what one obtains when considering the {\it integrated} avalanche distributions~\cite{durin2006role,perkovic1995avalanches} or distributions for quasiperiodic avalanche dynamics~\cite{papanikolaou2012quasi} which tend to exhibit pronounced bumps.

Second, 
our results provide an interesting perspective to understand the nature of criticality in these systems: Even if the external force is tuned by the driving mechanism to fluctuate in the immediate vicinity of the "global" critical force of the depinning phase transition, locally the interface and hence the localised avalanches via which the interface propagates behave as if the system was somewhat off-critical: either sub- or supercritical depending on the relative local height or the local curvature of the interface, such that the systems exhibit a mixture of behaviours around criticality. This should also imply non-zero predictability of the avalanche size, given information of the avalanche starting point~\cite{haavisto2023predicting,le2020correlations}. In contrast, the mean field system is truly critical in the sense that any large avalanche event, irrespective of its starting height and other attributes, follows the same distribution. Finally, one may expect that our conclusions extend to other spatially heterogeneous avalanching systems where bumps in avalanche distributions are often observed, e.g., lattice models of amorphous plasticity~\cite{budrikis2013avalanche} and sheared glasses~\cite{oyama2021unified}.

%

\appendix
\onecolumngrid
\begin{center}
    \textbf{\large End Matter}
\end{center}
\twocolumngrid

The above analysis shows how a bump emerges in the avalanche size distribution for interfaces with long-range and local elasticity due to an average excess contribution to the driving force arising from elasticity of the interface. Here we investigate the manner in which this effect depends on the parameter $k$ controlling the avalanche cutoff $s_0$. Of particular interest is how the effect manifests itself in the $k \rightarrow 0$ limit, where the avalanches become large and hence might in principle be less affected by their initiation point. To this end, we consider the LR model and the qEW equation with a large $L$ ($L=65536$ for the LR model and $L=32768$ for the qEW equation)
and study the 
behaviour of the $P(s)$ distributions for a broad range of $k$-values.  

\begin{figure}[t!]
    \centering
    \includegraphics[trim={0.7cm 3.9cm 1.3cm 4cm},clip,width=0.46\textwidth]{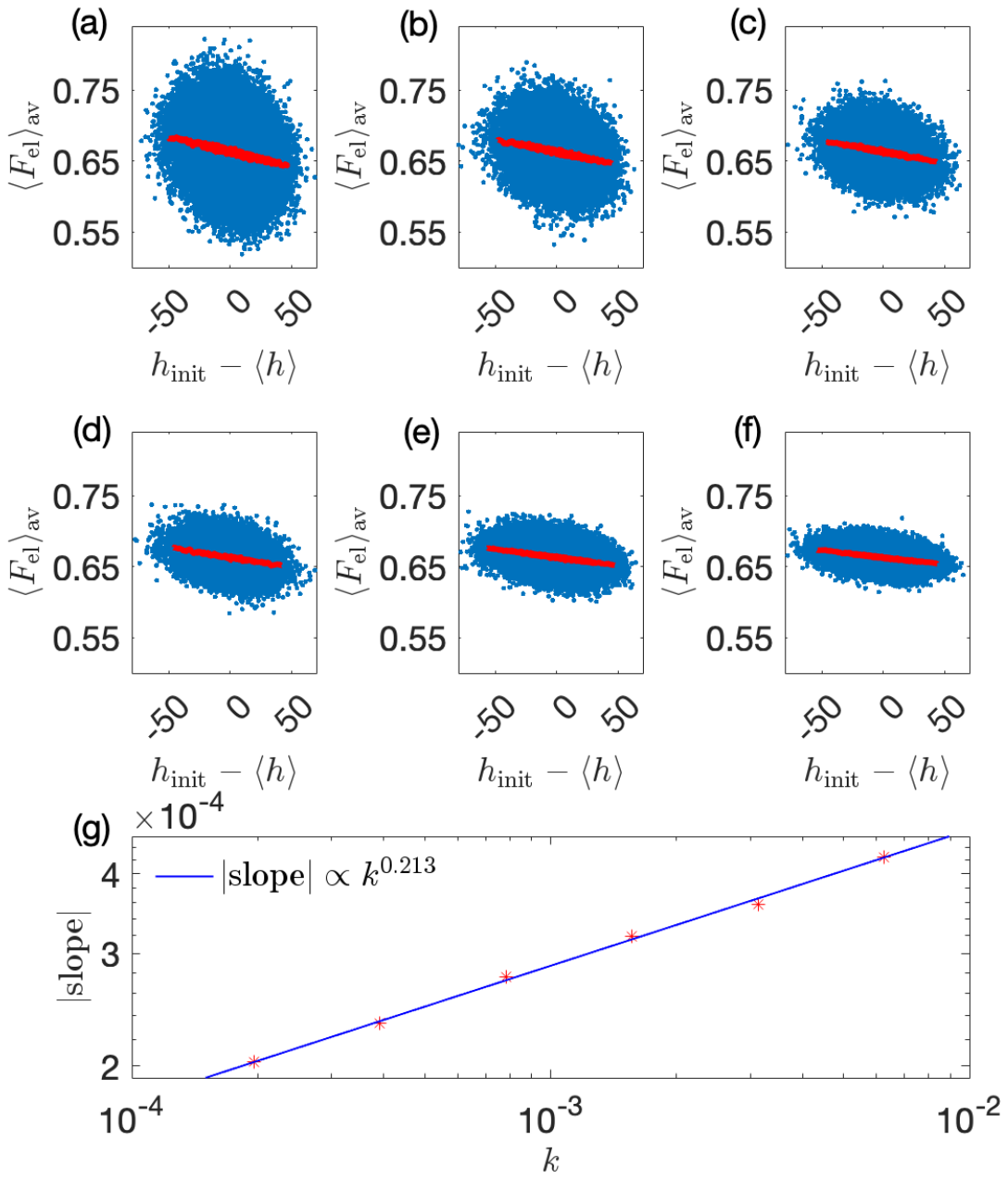}
    \caption{$\langle F_{\mathrm{el}} \rangle_{\mathrm{av}}$ of avalanches as a function of $h_\mathrm{init}-\langle h \rangle$ for the LR model with $L=65536$ for different values of $k$ and a $k$-dependent range $s \in [s_\mathrm{min}, s_\mathrm{max}]$ of avalanche sizes (blue dots). The running averages (red) are as in Fig.~\ref{fig:5}. In (a), $k=0.00625$, $s_\mathrm{min}=500$, and $s_\mathrm{max}=1000$. In (b), (c), ..., (f), the $k$-value is always half of that of the previous panel, and the previous $s_\mathrm{min}$ and $s_\mathrm{max}$-values are multiplied by a factor of $(1/2)^{-1/\sigma_k} = 1.6529$. Hence, in (f), $k=0.00009766$, $s_\mathrm{min}=6169$, and $s_\mathrm{max}=12339$. (g) shows the scaling of the absolute value of the slope of the running averages in (a)...(f), obtained from a linear fit, as a function of $k$. The solid line corresponds to a power law fit of the form of $|\mathrm{slope}| \propto k^{0.213}$.}
    \label{fig:6}
\end{figure}

We start by considering how the magnitude of the average elastic force acting on the avalanches in the LR model, $\langle F_\mathrm{el} \rangle_\mathrm{av}$ (the quantity considered earlier in Fig.~\ref{fig:5}), behaves as a function of $k$ and $s$. We vary $k$ between 0.00625 and 0.00009766 (with two subsequent $k$-values separated by a factor of 1/2), and consider for each $k$ a range of avalanche sizes that depends on $k$ so that $s \in [s_\mathrm{min}(k),s_\mathrm{max}(k)]$, with $s_\mathrm{min}(k) = 500(k/0.00625)^{-1/\sigma_k}$ and $s_\mathrm{max}(k) = 1000(k/0.00625)^{-1/\sigma_k}$. This choice is made in order to consider for each $k$ avalanches whose sizes have in relative terms the same distance from the cutoff scale $s_0 \propto k^{-1/\sigma_k}$, with $1/\sigma_k \approx 0.725$ for the LR model~\cite{laurson2010avalanches}. The resulting $\langle F_\mathrm{el} \rangle_\mathrm{av}$ vs $h_\mathrm{init}-\langle h \rangle$ plots are shown in Figs.~\ref{fig:6} (a) to (f), with the $k=0.00625$ case shown in (a), followed by reducing $k$ by a factor of 1/2 in each subsequent panel, until $k=0.00009766$ is reached in (f). We then measure the slopes of the running averages of $\langle F_\mathrm{el} \rangle_\mathrm{av}$ vs $h_\mathrm{init}-\langle h \rangle$ (shown in red) from Figs.~\ref{fig:6} (a) to (f). The absolute value of this slope is plotted on a log-log scale as a function of $k$ in Fig.~\ref{fig:6}(g), showing that the data is well-described by a power law $|\mathrm{slope}| \propto k^{0.213}$. This suggests that the "strength" of the dependence of $\langle F_\mathrm{el} \rangle_\mathrm{av}$ on $h_\mathrm{init}-\langle h \rangle$ decreases slowly with decreasing $k$ (and increasing $s$) and vanishes as $k \rightarrow 0$. We note that this is to be expected as $\langle F_\mathrm{el} \rangle_\mathrm{av}$ is averaged over the entire avalanche, and the elastic force tends to decrease during the avalanche propagation as the local height in the region of avalanche propagation increases faster than $\langle h \rangle$. This could then potentially imply that the mechanism giving rise to the bump in the $P(s)$ distributions discussed above would gradually disappear as larger and larger avalanches (or, smaller and smaller $k$-values) are considered.

\begin{figure}[ht!]
    \centering
    \includegraphics[trim={1.1cm 7.0cm 1.3cm 8cm},clip,width=0.48\textwidth]{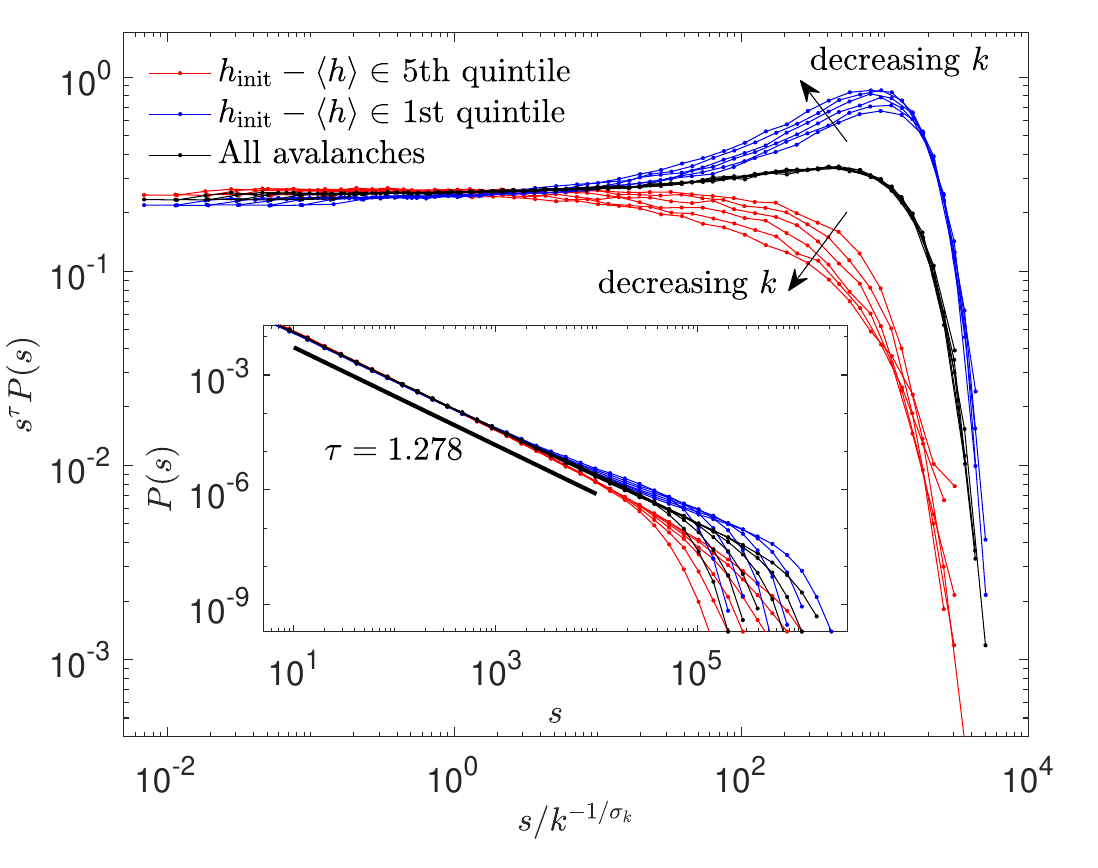}
    \caption{Inset: The avalanche size distributions $P(s)$ of the LR model with $L=65536$ for all avalanches (black) as well as avalanches with $h_\mathrm{init}-\langle h \rangle$ in the 1st and 5th quintile of the $P(h_\mathrm{init}-\langle h \rangle)$ distribution (red and blue, respectively). The different $P(s)$-distributions with the same color correspond to different $k$-values, with $k$ ranging from 0.00625 (smallest $s_0 \propto k^{-1/\sigma_k}$) to 0.00009766 (largest $s_0$). Main figure: The same distributions as in the inset, rescaled according to $P(s)=s^{-\tau}f(s/k^{-1/\sigma_k})$ with $\tau = 1.278$ and $1/\sigma_k = 0.725$. While a good data collapse is obtained for the full distributions, the distributions binned according to the quintiles of the $P(h_\mathrm{init}-\langle h \rangle)$ distribution exhibit $k$-dependent, non-universal scaling functions. Arrows indicate the direction of decreasing $k$.}
    \label{fig:7}
\end{figure}

\begin{figure}[ht!]
    \centering
    \includegraphics[trim={1.1cm 7.0cm 1.3cm 8cm},clip,width=0.48\textwidth]{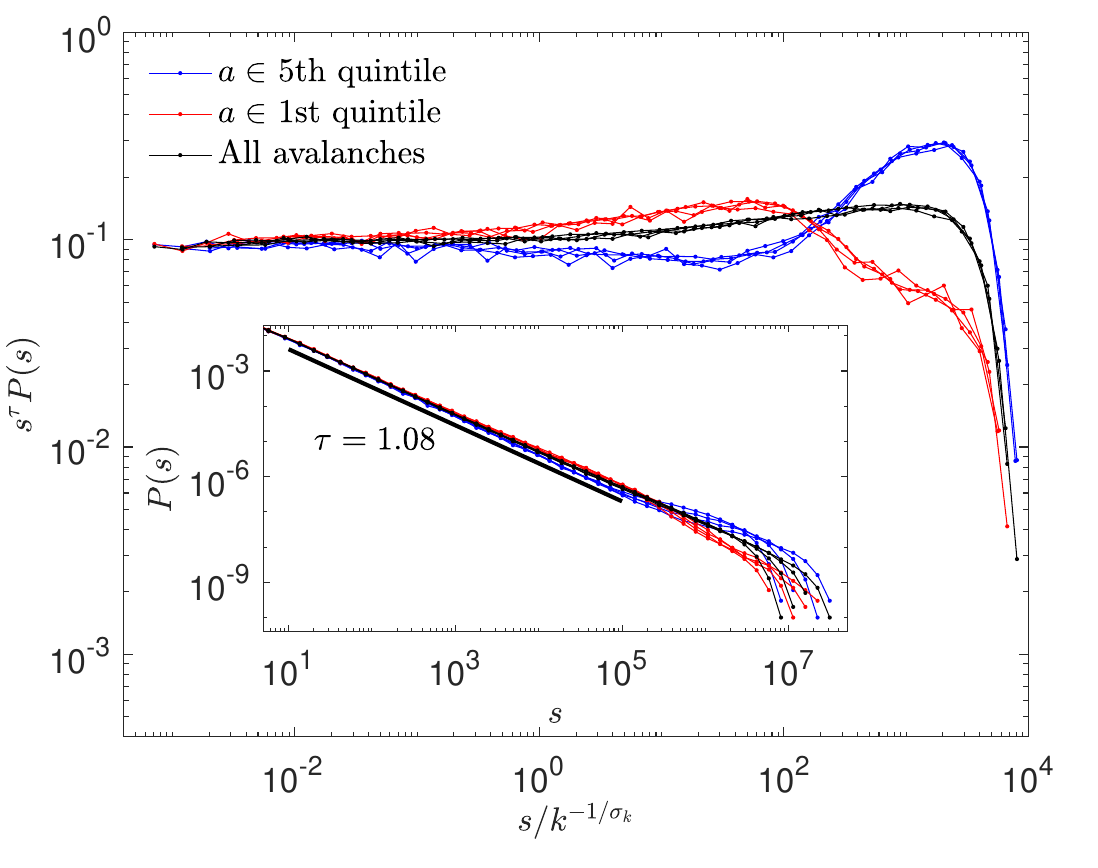}
    \caption{Inset: The avalanche size distributions $P(s)$ of the qEW equation with $L=32768$ for all avalanches (black) as well as avalanches with $a$ (extracted using the $k$-dependent fitting window, see main text) in the 1st and 5th quintile of the $P(a)$ distribution (blue and red, respectively). The different $P(s)$-distributions with the same color correspond to different $k$-values, with $k$ ranging from 0.0001 (smallest $s_0 \propto k^{-1/\sigma_k}$) to 0.0000125 (largest $s_0$). Main figure: The same distributions as in the inset, rescaled according to $P(s)=s^{-\tau}f(s/k^{-1/\sigma_k})$ with $\tau = 1.08$ and $1/\sigma_k = 0.75$. Good data collapses with distinct scaling functions are obtained for all three distributions.}
    \label{fig:8}
\end{figure}

The crucial point, however, is the question of how susceptible the avalanches are to this diminishing excess contribution to the driving force due to elasticity as $k$ is decreased. To estimate this, we proceed to analyze the avalanche size distributions $P(s)$ for different $k$-values. 
Inset of Fig.~\ref{fig:7} shows the full $P(s)$ distributions (black) for the same $k$-values considered above, as well as $P(s)$ of avalanches with $h_\mathrm{init}-\langle h \rangle$ in the 1st (bottom $20\%$) and 5th (top $20\%$) quintile of the $P(h_\mathrm{init}-\langle h \rangle)$ distribution (blue and red, respectively). Similarly to what was shown above for a fixed $k$, $P(s)$ for avalanches with $h_\mathrm{init}-\langle h \rangle$ in the 1st quintile and in the 5th quintile clearly exhibit supercritical and sub-critical characteristics, respectively, for all $k$-values considered. For a more detailed look, we attempt to collapse the distributions for different $k$ according to the scaling form $P(s,k)=s^{-\tau}f(s/k^{-1/\sigma_k})$ (main figure of Fig.~\ref{fig:7}). The good data collapse obtained for the full distributions (black) shows that they exhibit the well-known behaviour of following the above scaling form with a universal scaling function $f(x)$ with a bump. The distributions binned according to the quintiles of the $P(h_\mathrm{init}-\langle h \rangle)$ distribution, however, exhibit subtle deviations from the scaling form applicable to the full $P(s)$. First, the bump of the ($k$-dependent) $f(x)$ for avalanches with $h_\mathrm{init}-\langle h \rangle$ in the 1st quintile becomes more prominent as $k$ decreases, i.e., the rescaled distributions exhibit a larger maximum at the bump for smaller $k$. Second, the opposite effect can be seen for avalanches with larger $h_\mathrm{init}-\langle h \rangle$ -values: Avalanches with $h_\mathrm{init}-\langle h \rangle$ in the 5th quintile become increasingly sub-critical-like as $k$ is decreased towards zero (the arrows in the main figure of Fig.~\ref{fig:7} indicate the direction in which $k$ decreases). Thus, the gap between the super- and sub-critical avalanches, obtained at the two extremes of the $P(h_\mathrm{init}-\langle h \rangle)$ distribution, actually widens as $k$ decreases, implying that the avalanche initiation point remains relevant even in the limit of small $k$ and large avalanches. We note that this is true even if the difference between $\langle F_\mathrm{el} \rangle_\mathrm{av}$ at the two extremes of the $P(h_{\mathrm{init}}-\langle h \rangle)$ distribution, characterized by the slope in Fig.~\ref{fig:6}(g), decreases with decreasing $k$. This shows that the tendency of the system to exhibit increased susceptibility to perturbations closer to the critical point outweighs the slowly decreasing strength of the excess contribution of elasticity to the driving force as $k$ is decreased towards zero. 

Finally, we repeat the above analysis for the qEW equation, focusing on the $P(s)$ distributions binned according to the $a$-parameter, estimated here by considering a fitting window with a $k$-dependent length $l_x$ around the avalanche starting point. $l_x$ is chosen to scale as the lateral length of the cutoff avalanches, such that $l_x \propto k^{(-1/\sigma_k)/(1+\zeta)}$, where $\zeta$ is the roughness exponent. $\zeta \approx 1.25$~\cite{kim2006depinning} and $1/\sigma_k = \nu_k (d+\zeta) = 1/3 (1+\zeta) \approx 0.75$~\cite{durin2000scaling} then implies $l \propto k^{-0.33}$. In Fig.~\ref{fig:8}, we take the largest $k$-value of 0.0001 to correspond to $l_x = 1000$, which then implies $l_x = 1287$, 1617 and 2033 for $k=0.00005$, 0.000025 and 0.0000125, respectively. This choice results in good data collapses for all three cases shown in Fig.~\ref{fig:8}, i.e., the 1st and 5th quintiles of the $P(a)$ distribution (corresponding to super- and subcritical avalanches, respectively), as well as for the full $P(s)$, suggesting that the local curvature around the avalanche starting point remains relevant as $k \rightarrow 0$. The analysis also shows that the quantity that was chosen to be used for binning is here the "correct" one in the sense that it results in data collapses also for the super- and sub-critical avalanches. At the same time, for each $k$ and just like for the LR model, the average over all the quintile-binned distributions equals the full $P(s)$ distribution, which exhibits a universal scaling function $f(x)$ with a bump.

\end{document}